\documentclass[twocolumn,showpacs,preprintnumbers,amsmath,amssymb]{revtex4}

\usepackage{graphicx}
\usepackage{dcolumn}
\usepackage{bm}
\usepackage{multirow}
\usepackage{color}
\usepackage{ulem}

\begin{document}

\title{Characterizing top gated bilayer graphene interaction with its environment by Raman spectroscopy}

\author{D. L. Mafra$^{1,*}$, P. Gava$^{2,*}$, L. M. Malard$^1$,
R. S. Borges$^3$, G. G. Silva$^3$, J. A. Leon$^1$,
F. Plentz$^1$, F. Mauri$^2$, M. A. Pimenta$^1$}
\address{$^1$Departamento de F\'{\i}sica, Universidade Federal de Minas Gerais, 30123-970, Belo Horizonte, Brazil.\\
$^2$IMPMC, Universitè Paris 6 et 7, CNRS, IPGP, Paris, France.\\
$^3$Departamento de Qu\'{\i}mica, Universidade Federal de Minas
Gerais, 30123-970, Belo Horizonte, Brazil.\\
$^*$ These authors contributed equally to this work.}

\begin{abstract}

In this work we study the behavior of the optical phonon modes in
 bilayer graphene devices by applying top gate voltage, using Raman scattering.
 We observe the splitting of the Raman G band as we tune the Fermi level of
 the sample, which is explained in terms of mixing of the Raman ($E_g$) and infrared
 ($E_u$) phonon modes, due to  different doping in the two layers. We theoretically analyze
 our data in terms of the bilayer graphene phonon self-energy which includes
 non-homogeneous charge carrier doping between the graphene layers. We show
 that the comparison between the experiment and theoretical model not only gives
 information about the total charge concentration in the bilayer
 graphene device, but also allows to separately quantify the amount of unintentional charge coming
from the top and the bottom of the system, and therefore to characterize the interaction of
bilayer graphene with its surrounding environment.

\end{abstract}

\maketitle


Bilayer graphene has attracted a lot of attention recently because
of its special low energy electronic dispersion, in which a tunable
band gap can be opened by application of a transverse electric
field \cite{mccann,netoreview,ohta,neto07biased,oostinga07,wang2009,
mak2009,kuzmenko-bandgap}. Such device is desirable for low energy
photo-emitters and detectors possessing a high tunability by the
control of charge concentrations on the graphene layers. Recent
experimental demonstration of this tunable band gap in bilayer
graphene was based on the absorption measurements in the infrared
region \cite{wang2009,mak2009,kuzmenko-bandgap} or by electric
transport measurements\cite{neto07biased,oostinga07}. However the
tunable band gap bilayer graphene device operation can be greatly
influenced by the surrounding environment. Typically, unintentional doping charges coming from the top and the bottom
of the system can accumulate on bilayer graphene, giving rise to
an unintentional electric field which determines a
non-homogeneous doping between the layers and the opening of a band gap
in the band structure, without any applied electric field
\cite{paola}. In this work we use Raman spectroscopy to
monitor the unintentional charge coming from the top and the bottom of the system,
which gives information on the electrostatic environment of the sample and
which helps to characterize the bilayer devices for further
applications.

The band gap opening and tunability in bilayer graphene is based on
the application of an electric field $E$ perpendicular to the
layers, given by $E=(n_{top}-n_{bot})|e|/(2\epsilon_0)$ where
$n_{top}$ and $n_{bot}$ are the charge carriers coming from the top
and the bottom of bilayer graphene, respectively. Raman spectroscopy
has already shown to be a fast and non-destructive tool to
characterize graphene samples \cite{ferrari,malard09} and doping
effects \cite{pisana07,yan07,das08,das09,bruna}, however no
carefully analysis has been done to demonstrate the effect of
non-homogeneous doping in bilayer graphene devices. Recent
theoretical calculations made by Gava\,\emph{et al.} \cite{paola}
suggest that from the analysis of the Raman spectra of gated bilayer
graphene it is possible to quantitatively identify the amount of
non-intentional charges coming from the atmosphere and from the
substrate and to characterize the electrostatic environment of
few-layers graphene.  In this work we study the dependence of the G
band of bilayer graphene on the gate voltage. From the direct
comparison between the experimental and the theoretically simulated
Raman spectra, and from the analysis of the positions, full width at
half maximum (FWHM) and relative intensities of the two Raman peaks
as a function of the electron concentration, we were able to
estimate the charge unintentionally accumulated on the device from
the environment.

\begin{figure}
\includegraphics [scale=0.87]{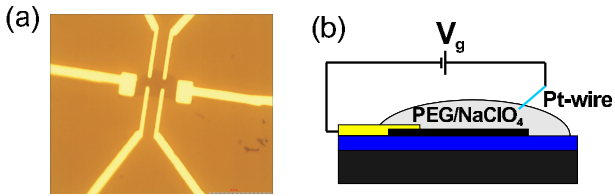}
\caption{\label{Fig1} (a) Optical microscope image of the graphene
sample before application of the polymer electrolyte. (b) Schematic
illustration of the device and the experimental setup.}
\end{figure}

Fig.\,\,\ref{Fig1}(a) shows the bilayer-graphene field-effect
transistor (FET) used in the experiment. Graphene samples were
produced by micro-mechanical cleavage of graphite and deposited on
Si covered with 300nm of SiO$_2$. Top gating was achieved by using a
polymer electrolyte consisting of polyethylene glycol (PEG) and
NaClO$_4$ with ratio concentration of 1:0.25, and the gate voltage was applied between a gold electrode
in contact with the graphene layer and a platinum wire electrode
inserted in the electrolyte (see the schematic setup in
Fig.\,\,\ref{Fig1}(b)). The contacts were made by optical
lithography. The Raman measurements were done in the back scattering
configuration at room temperature using a triple monochromator
spectrometer (DILOR XY) using 2.41\,\,eV as excitation laser energy.
The spot size of the laser was $\sim$1\,$\mu$m using a 80\,$\times$
objective and the laser power was kept at 1.4\,\,mW.


The  interface between graphene and polymer electrolyte has been
shown experimentally to behave like a double layer capacitor
of thickness in the order of nanometers \cite{bard}. Therefore, the
geometric capacitance of the electrolyte is very high compared to
bottom gate devices where the thickness of the dielectric is much
larger (typically 300 nm).

Capacitance measurements of the polymeric electrolyte used in the
experiment were performed by Impedance Spectroscopy with frequency
analyzer AUTOLAB PGSTAT30 by using a symmetrical cell with two Au
electrodes and a polymer electrolyte layer. The impedance
measurements for the electrolytes were carried out with frequency
ranging from 50\,\,kHz to 0.5\,\,Hz at 0\,\,V with 10\,\,mV
amplitude, and we obtained the value of 1.5 $\times$ 10$^{-6}$ F
cm$^{-2}$. However, the shape and the thickness of the electrical
double layer depends on the specific surface at contact with the
electrolyte. Therefore, the measured value of C$_G$ only gives the
order of magnitude of the geometric capacitance of the electrolyte
in contact with bilayer graphene.

In the case of bilayer graphene with Bernal AB layer stacking, both
the electronic and phonon bands split into two components with
special symmetries \cite{tggraphene}. The E$_{2g}$ phonon mode of
monolayer graphene splits into two components, associated with the
symmetric (S) and anti-symmetric (AS) displacements of the atoms in
the two layers with respect to inversion symmetry
\cite{malardprl08}. The S and AS modes belong to the two double
degenerated representation E$_g$ and E$_u$, respectively
\cite{paola,ando09,malardprl08,yan09}. The E$_u$ mode is not Raman
active and, therefore, the G band of isolated bilayer graphene is
composed of only one peak. However, when the two layers of bilayer
graphene have different charge carrier concentration, induced by the
application of an external gate voltage, the inversion symmetry of
bilayer graphene is broken, lowering the symmetry of the system. As
a consequence of the induced asymmetry between the two layers, the
two S and AS modes are mixed, the two new eigen-modes have the Raman
active S component, and therefore two peaks are observed in the G
band of bilayer graphene \cite{ando09,paola,malardprl08,yan09}.

\begin{figure}
\includegraphics [scale=0.5]{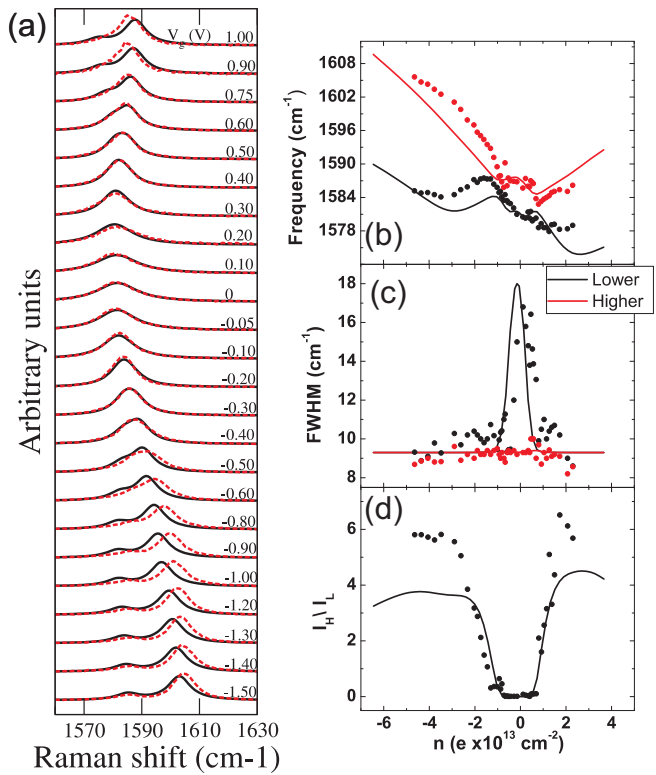}
\caption{\label{fig2} (a) Comparison between experimental (red dashed line)
and theoretical (black continuous line) spectra for different V$_g$ as obtained
from the direct fit of spectra; (b-c-d) Frequency $\omega_{h/l}$,
FWHM $\Gamma_{h/l}$ and ratio of I$_{h/l}$ as obtained fitting the
experimental spectra with two Lorentzians and compared with theory.
The dots are the experimental data and the full curves are the
theoretical calculations.}
\end{figure}

In Fig.\,\,\ref{fig2}(a) (red dashed curves) we show the experimental Raman
spectra taken with the application of top gate voltage (V$_g$) from
-1.50 to 1.00\,\,V. The G band splitting into two components G$_h$
and G$_l$ (higher and lower frequency peak, respectively) can be
clearly observed for V$_g$ below -0.6\,\,V. In order to compare the
experimental spectra and theoretical calculations, we converted
V$_g$ into $n$ using the expression
$\beta V_g = (n - n_0) e$
where the total capacitance $\beta$, which includes the quantum
capacitance (C$_Q$) and geometrical capacitance (C$_G$)
\cite{das08}, and the intrinsic doping at the zero gate n$_0$ are
used as fitting parameters. Moreover, by the comparison between
experimental and theoretical results we can estimate the charges
unintentionally adsorbed, at zero gate, from the top and bottom of
the device, $n_{top}^{0}$ and $n_{bot}^{0}$ respectively. These
quantities are related to n$_0$ by $n_0=n_{top}^{0}+n_{bot}^{0}$,
and therefore we only used n$_{bot}^{0}$ as additional fitting
parameter. Finally, the theoretical FWHM $\Gamma^{th}$ calculated as
a function of n and n$_{bot}^{0}$ is given by electron-phonon and
an-harmonic phonon-phonon interaction \cite{paola}. Therefore, in
order to take into account other factors determining a finite
lifetime and neglected in the calculations, we used in the fitting
procedure a parameter $\Gamma^0$, independent on the total charge
$n$ and equal for the two peaks, related to the total FWHM by
$\Gamma$ = $\Gamma^{th}$ + $\Gamma^0$.

In Fig.\ref{fig2}(a) we show the comparison between the experimental
spectra (red dashed curve) and theoretical one (black continuous curve), obtained
using the parameters discussed below, for different V$_g$. The fit
is performed computing the square of the difference between the
experimental and theoretical spectra, averaged over all the measured
Raman range and over different V$_g$. We considered V$_g$ in the
range of $\pm$0.5 V. This choice is motivated by the fact that for
large values of V$_g$ the linear relation between gate voltage and
charge could be modified, and charges from the electrolyte could
accumulate on the bottom of the sample, making the fit results less
reliable. The theoretical spectra is obtained as the sum of two
Lorentzians, for the G$_h$ and G$_l$ frequency peaks. The two
Lorentzians are centered in $\omega_{h/l}$, with FWHMs
$\Gamma_{h/l}$ = $\Gamma^{th}_{h/l}$ + $\Gamma^0$, and with area
I$_{h/l}$, where $\omega_{h/l}$, $\Gamma^{th}_{h/l}$, and I$_{h/l}$
are computed as a function of $n$ and n$_{bot}^{0}$. In particular,
we used two different parameters, $\beta^+$ and $\beta^-$, for
positive and negative $V_g$, which induce positive and negative $n$
(i.e. electron and hole doping charge), respectively. The values for
the parameters used in the fit (i.e. $\beta^+$, $\beta^-$, n$_0$,
n$_{bot}^{0}$, and $\Gamma^0$) are varied in uniform and dense
grids. The best fit is obtained for $\beta^+$ = 3.7 and $\beta^- =
4.6 \times 10^{-6}$ F cm$^{-2}$, n$_0 = -0.15 \times 10^{13}$
cm$^{-2}$, n$_{bot}^{0}$ = 0.0, and $\Gamma^0$ = 7.5 cm$^{-1}$.
Notice that the agreement between the experimental and simulated
spectra is excellent in the range -0.5 to 0.5\,\,V. The slight
shifts out of this range can be ascribed to a possible hysteresis in
the experiment, and to the fact that we did not consider in
our model the  expected jump of the quantum capacitance
($C_Q$) when, increasing (decreasing) the Fermi
level, we reach the second conduction band (penultimate valence band)
in bilayer graphene \cite{das08}.
The different values of $\beta^+$ and
$\beta^-$ can be ascribed to the different mobilities of the
positive ($Na^+$) and negative ($ClO_4^-$) ions.

All the G band experimental spectra of Fig.\,\ref{fig2}(a) were
fitted using two Lorentzians in order to extract the frequencies and
full width at half maximum (FWHM) of two components G$_h$ and G$_l$,
as well as the relative Raman intensity, i.e., the ratio between the
Raman intensities of the modes with highest frequency (I$_h$) and
lowest frequency (I$_l$). Figs.\,\,\ref{fig2}(b) and (c) shows,
respectively, the dependence of the G$_h$ (red dots) and G$_l$
(black dots) frequencies and FWHM $\Gamma$ as a function of the
electron concentration. The full lines are the theoretical
calculations. The dependence of the frequency of the G$_{h}$ and
G$_{l}$ Raman peaks (Fig. \ref{fig2}(b)) is well described by the
calculation of the phonon self energy as a function of charge
concentration. The distinct behaviors of the G$_{h}$ and G$_{l}$ is
qualitatively explained by the distinct electron-phonon couplings of
these modes. While G$_{h}$ blue-shifts with charge carrier
concentration due to inter-band transitions, the G$_{l}$ mode
redshifts due to intra-band transitions when the E$_{F}$ is changed.
For the FWHM dependence, while $\Gamma_{l}$ does not change with
$n$, the $\Gamma_{h}$ is maximum near $n=0$ and minimum for values
of $n$ corresponding to values of $E_F$ larger than half of the
phonon energy, as has been observed before in both mono and bilayer
graphene \cite{pisana07, yan07, das08}. It's worth to mention that
the scattered data points for the FHWM are mostly caused by charge
carrier fluctuation during the measurement, where a hysteresis of
the charge neutrality point is found by sweeping the gate voltage up
and down. In Fig.\,\,\ref{fig2}(d) we plot the dependence of the
I$_h$/I$_l$ as a function of $n$. The quantity I$_h$/I$_l$ shows a
minimum value between n$\sim$ -1 to 1 $\times$ $e$ 10$^{13}$
cm$^{-2}$, and increases more strongly for positive carrier
concentration.


In summary, a detailed analysis of the G band of top gated bilayer
graphene is presented. We observed that, unlike in the unbiased case
where the G Raman band is composed by only one peak, the gate
voltage breaks the inversion symmetry and the G band splits in two
modes, that are combinations of the symmetric and anti-symmetric
modes of the unbiased bilayer graphene. We analyze the dependence of
the frequency and the relative intensities of the peaks with higher
and lower frequency as a function of the electron concentration and
we compared the experimental results with theoretical calculations.
From this comparison, we could estimate the unintentional carrier
concentration adsorbed on the device, at zero gate, from the
substrate, $n^{0}_{bot}$, and from the electrolyte, $n^{0}_{top}$,
and we found $n^{0}_{bot}=0.0$ and $n^{0}_{top}= -0.15 \times
10^{13}$\,\,cm$^{-2}$, showing that Raman spectroscopy is a powerful
technique to study the electrostatic environment of graphene.


We would like to acknowledge Nacional de Grafite (Brazil) for
providing us the graphite samples. D.L.M. and L.M.M. acknowledges
the support from the Brazilian Agency CNPq.

\end{document}